\documentclass[aps,prd,nofootinbib]{revtex4-2}
\usepackage[colorlinks=true, pdfstartview=FitV, linkcolor=blue, citecolor=red, urlcolor=magenta]{hyperref}
\usepackage{graphicx}
\usepackage{latexsym}
\usepackage{amsmath}
\usepackage{amsfonts}
\usepackage{amssymb}
\usepackage{booktabs} % Required for \toprule, \midrule, and \bottomrule
\usepackage{verbatim}
\usepackage[dvipsnames]{xcolor}
\usepackage[most]{tcolorbox}
\newtcolorbox{revnote}[1][]{colback=orange!6,colframe=orange!70!black,
  fonttitle=\bfseries\small,fontupper=\small,title={Revision note #1},
  left=4pt,right=4pt,top=2pt,bottom=2pt,breakable}

\newcommand{\be}{\begin{equation}}
\newcommand{\ee}{\end{equation}}
\newcommand{\bea}{\begin{eqnarray}}
\newcommand{\eea}{\end{eqnarray}}

\newcommand{\ben}{\begin{eqnarray}}
\newcommand{\een}{\end{eqnarray}}
\newcommand{\eps}{\varepsilon}

\begin{document}

\title{Radiative stability of species-separated Kaluza--Klein spectra\\
in warped geometry}

%%%%%%%%%%%%%%%%%%%%%%%%%%%%%%%%%%%%%%%%%%%%%%%%%%%%%%%%%%

\author{M. A. Anacleto$^{1,2}$}
\email{anacleto@df.ufcg.edu.br}

\author{F. A. Brito$^{1,2}$}
\email{fabrito@df.ufcg.edu.br}

\author{E. Passos$^{1,2}$}
\email{passos@df.ufcg.edu.br}

\author{Joao R. L. Santos$^{1,2}$}
\email{joaorafael@df.ufcg.edu.br}

\author{J. A. V. Campos$^{1}$}
\email{joseandrecampos@gmail.com}

\author{I. E. L. M. Santana$^{1}$}
\email{isaque.ufcg@gmail.com}

\affiliation{$^{1}$Departamento de F\'{i}sica, Universidade Federal de Campina Grande, Caixa Postal 10071, 58429-900, Campina Grande, Para\'{\i}ba, Brazil.}

\affiliation{$^{2}$Unidade Acad\^emica de Matem\'atica, Universidade Federal de Campina Grande,\\ 58429-970, Campina Grande, Para\'{\i}ba, Brazil.}
%%%%%%%%%%%%%%%%%%%%%%%%%%%%%%%%%%%%%%%%%%%%%%%%%%%%%%%%%%

\begin{abstract}
Species-dependent fifth-direction lightcones, generated by scalar-gradient operators, can separate gauge and fermion Kaluza--Klein (KK) scales in a warped extra dimension, as recently shown by Randall and Riojas. In this work, we study whether this separation survives quantum corrections. Quark loops can only move the gluon speed within the interval spanned by the bare speeds. Gluon exchange pulls the quark speed toward the gluon speed, with a one-loop coefficient strongly enhanced by fifth-component exchange when the gluon is the slow species. We set the size of this effect using the strong-coupling estimates for a light-gluon benchmark, where a $5.5$~TeV gluon sits below quark partners at $10.74$~TeV. Within the effective theory, the Kaluza--Klein (KK) modes then correct the quark speed at order one. That correction is several times larger than the mass ordering can absorb, so the light-gluon ordering has no perturbative protection. The reverse benchmark, with slow quarks and a heavy gluon, keeps its ordering roughly ten times more robust.
\end{abstract}
\maketitle
% ==================================================================
\section{Introduction}
% ==================================================================

Warped extra dimensions explain the hierarchies of the Standard Model through the localization of fields along the fifth dimension~\cite{RS1,GP,GN}. The same geometry that sets the Yukawa couplings also fixes the Kaluza--Klein (KK) spectrum, and in the simplest constructions gauge and fermion resonances share a common scale. Flavor constraints then push that scale beyond the reach of colliders. Randall and Riojas~\cite{RR} recently showed that this link can be broken without changing the zero-mode flavor structure. In their approach, the scalar-gradient operators give each species its own propagation speed along the fifth dimension, while the four-dimensional light cone remains common. Therefore, a field with a reduced fifth-direction speed sees a deeper effective throat, and its whole KK tower is lowered by the same factor. It is relevant to point out that related species-dependent KK scales had been studied in bulk Lorentz-violating and aether compactifications, as we can see in  ~\cite{CarrollTam,Rizzo2005,Rizzo2010,Dubovsky,Oliveira2020}.

Ref.~\cite{RR} presents two benchmarks on a stabilized background with $kL=28$. In Model~A, the right-handed quarks are slow, $v_S\simeq0.2$, and their partners lie at $6.6$--$12$~TeV, well below a $30$~TeV KK gluon. In Model~B, only the gluon is slowed, $u_g=0.44$. This places a narrow, top-philic KK gluon at $5.50$~TeV, below quark partners that begin at $10.74$~TeV. Model~B is the more striking of the two for collider physics, and the more delicate, since its mass ordering rests on a single speed ratio. The authors listed the ultraviolet origin and the radiative stability of the unequal speeds among the open questions of their construction. This paper addresses the second one.

The question is natural. Speed differences between interacting species are not protected by any symmetry. In four dimensions they flow logarithmically toward a common value~\cite{ChadhaNielsen,AnberDonoghue}, and power-law divergences can turn small bare Lorentz violation into large effects as shown by Collins et al. ~\cite{Collins2004}. In five dimensions the relevant loops are linearly divergent. Their finite parts are matching data, but the contribution of the KK modes that the effective theory itself describes grows linearly with the number of modes, as in the power-law running of extra-dimensional gauge couplings~\cite{DDG}. It is relevant to highlight that the radiative corrections to KK masses have been computed for universal extra dimensions~\cite{CMS,Freitas2018a,Freitas2018b} and, in a setting closer to the present one, for five-dimensional QED with an aether background~\cite{Oliveira2020}. In that work the divergent local part of the fermion loop was subtracted, and the finite compactification effect was computed. For the stability of a species-separated spectrum, it is precisely the local part that matters.

Our analysis proceeds in three steps. We first observe that, on the background of Ref.~\cite{RR}, the gauge operator is exactly an anisotropic-metric kinetic term, and that the quarks of Model~B are five-dimensionally relativistic. Quark loops can then only generate gauge operators aligned with the relativistic light cone. As a result, the renormalized gluon speed is a weighted average of the bare gluon and quark speeds: it can approach the quark speed but never exceed it. Quark loops alone would make the gluon heavier than the partner only if the induced coefficient exceeded about twice the tree-level one. Second, we compute the opposite effect, the renormalization of the quark speed by gluon exchange. Here the slow gluon matters: exchange of its fifth component is weighted by the square of the speed ratio, and the quark speed responds as $\delta\ln u_q\simeq-6.1\,c'$, where $c'$ is the one-loop coefficient. This is about four times the size of an ordinary wave-function renormalization. We have confirmed the coefficients by an independent Euclidean computation. Third, we turn these results into a criterion for the mass ordering through a spectral comparison argument, which is exact when the induced coefficients are constant.

To decide whether Model~B satisfies the criterion, we need the size of $c'$. Ref.~\cite{RR} quotes its strong-coupling estimate at several energies and for three gluon speeds. These values grow linearly with energy, scale exactly as the inverse gluon speed, and agree to $4\%$ with standard naive dimensional analysis measured in the gluon's light cone. With this calibration, the KK modes between the quark partner and the strong-coupling scale give $c'\simeq0.4$. That is an order-one renormalization of the quark speed, and it exceeds the value tolerated by the ordering by a factor of about $7$ for the Model~B quark content
($2$--$11$ over the plausible range of inputs). The light-gluon ordering is therefore not perturbatively protected: its fate depends on ultraviolet matching data and on effects beyond one loop. The regulator-independent content of this statement lies in the thresholds of the two gluon KK levels inside the window, at $12.47$ and $19.47$~TeV.

Applied to Model~A, the same framework reverses the roles. With slow quarks and a relativistic gluon, the fifth-component enhancement disappears and the sensitivity moves to the gauge sector. The ordering then survives up to an order-one renormalization, about twelve times more than in Model~B. The radiative robustness of species-separated spectra is thus governed by which species is slow.

The work is organized as follows. Section~\ref{sec:setup} reviews the operators of Ref.~\cite{RR} and sets up the spectral problem. Section~\ref{sec:gauge} derives the alignment of the induced gauge operators, and Sec.~\ref{sec:fermion} computes the quark-speed renormalization. Section~\ref{sec:criterion} combines them into a criterion for the ordering, which Sec.~\ref{sec:modelB} applies to Model~B and Sec.~\ref{sec:modelA} to Model~A. Section~\ref{sec:conclusions} summarizes the results and their limitations. Technical details are collected in the appendices. All statements concern the real KK spectrum, not the complex resonance poles.

% ==================================================================
\section{Species-dependent lightcones}\label{sec:setup}
% ==================================================================

Following Ref.~\cite{RR}, the species-dependent propagation is generated by the operators
\begin{equation}
\Delta\mathcal L_\Psi=-\frac{i\alpha_{AB}}{2\Lambda_0^5}\,\partial_M\phi^A\partial^N\phi^B\,
\bar\Psi\Gamma^M\overset{\leftrightarrow}{D}_N\Psi,\qquad
\Delta\mathcal L_g=\frac{\eta_{AB}}{2g_5^2\Lambda_0^5}\,\partial^M\phi^A\partial^N\phi^B\,
F^a_{MP}F^{a\,P}_{N},
\label{eq:RRops}
\end{equation}
where $\phi^A$ and $\phi^B$ are bulk scalars and $\Lambda_0$ is the operator scale. On a background with constant gradient $\partial_M\phi^A=\nu^A\delta^5_M$, they set the fifth-direction speeds
$v_\Psi=1-\alpha\nu^2/\Lambda_0^5$ and $u_g^2=1-\eta_g\nu^2/\Lambda_0^5$. Model~B has $v_\Psi=1$
for all quarks and $u_g=0.44$, which corresponds to $\eta_g\nu^2/\Lambda_0^5=0.8064$.

Describing each species by its own effective metric, and taking $n^M$ as the unit normal to the branes, we define
\begin{equation}
G_s^{MN}=g^{MN}+(u_s^2-1)\,n^Mn^N,\qquad \sqrt{|G_s|}=\frac{\sqrt{|g|}}{u_s}.
\label{eq:Gs}
\end{equation}
For an antisymmetric field strength, the term quartic in $n$ vanishes and
\begin{equation}
G_g^{MP}G_g^{NQ}F_{MN}F_{PQ}=F_{MN}F^{MN}+2(u_g^2-1)\,n^Mn^PF_{MN}F_P^{\ N}
\label{eq:identity}
\end{equation}
holds exactly. The ordinary gauge kinetic term plus $\Delta\mathcal L_g$ is therefore
\begin{equation}
-\frac{1}{4g_5^2}\sqrt{|g|}\,G_g^{MP}G_g^{NQ}\,{\rm tr}F_{MN}F_{PQ},
\end{equation}
which is the form written in Ref.~\cite{RR}, where fermions are similarly described by a species-dependent inverse vielbein, $\tilde e_{s\,a}{}^M=e_a{}^M+(v_s-1)n_an^M$, corresponding to the vielbein of $G_s$. We use the normalization of $g_5$ fixed by Eq.~\eqref{eq:identity} throughout. These statements characterize the two-derivative background operators and do not exclude other invariants involving scalar fluctuations or the boundaries.

On the metric $ds^2=e^{-2A(y)}\eta_{\mu\nu}dx^\mu dx^\nu-dy^2$, $0\le y\le L$, the transverse
gluon modes obey
\begin{equation}
-\partial_y\big[p(y)\,\partial_ya_n\big]=m_{g,n}^2\,w(y)\,a_n,\qquad p=u_g^2e^{-2A},\quad w=1 .
\label{eq:SL}
\end{equation}
Brane-localized kinetic terms with weights $r_i\ge0$ contribute $\sum_i r_i f(y_i)h(y_i)$ to the norm. The masses are the eigenvalues of the pair of quadratic forms $a[f,h]=\int p\,f'h'$ and $b[f,h]=\int w\,fh+\sum_ir_if(i)h(i)$. This formulation allows the min--max principle~\cite{Kato} to map bounds on the forms directly into bounds on the masses. If radiative corrections change the forms so that
$a_-a_0\le a\le a_+a_0$ and $b_-b_0\le b\le b_+b_0$, then
\begin{equation}
\frac{a_-}{b_+}\,m_{g,n}^{(0)2}\le m_{g,n}^2\le\frac{a_+}{b_-}\,m_{g,n}^{(0)2}\,,
\label{eq:comparison}
\end{equation}
for every massive level. Two consequences will be used repeatedly. First, if $p$ and $w$ are rescaled by constants, the profiles do not change and the masses scale exactly by the square root of the ratio. In particular $m_{g,n}\propto u_g$, as noted in Ref.~\cite{RR}. Second, positive brane kinetic terms enlarge $b$ at fixed $a$, so they can only lower the gauge masses~\cite{DHR,CPTW}. Since the issue for Model~B is a gluon that becomes too heavy, induced positive brane terms never threaten the ordering.

The first consequence underscores the necessity of a full radiative analysis. Holding the geometry and the fermions fixed, the gluon mass is $m_{g,1}=5.50\,(u_g/0.44)$~TeV. The ordering would be lost at $u_g=0.859$, i.e.\ at $\eta_g\nu^2/\Lambda_0^5=0.262$. The tree-level benchmark does not determine whether quantum corrections pull the effective speeds toward that point, and a fermion-loop correction computed in a different theory, such as the one of Ref.~\cite{Oliveira2020}, cannot settle the issue. We therefore compute the corrections within the Model~B effective theory itself.

If the gluon and quark masses change to $m_{g,1}=5.50\,G$ and $m_{q,1}=10.74\,F$~TeV, the ordering $m_{q,1}>m_{g,1}$ holds as long as
\begin{equation}
G<1.952727\,F .
\label{eq:cert}
\end{equation}
Equivalently, writing $G=1+\epsilon_g$ and $F=1+\epsilon_q$, the condition is $\epsilon_g-1.952727\,\epsilon_q<0.952727$. A common bound $|\epsilon_{g,q}|<0.3226$ is therefore sufficient. In the next two sections, we present the details about the computation of $G$ and $F$.

% % ==================================================================
\section{Gauge sector: aligned operators and convexity}\label{sec:gauge}
% ==================================================================

At one loop, the bulk gauge kinetic term receives contributions from gluon, ghost, and $A_5$ loops alongside quark loops. In Model~B the quarks have $v_\Psi=1$, meaning their determinant matches that of an ordinary Lorentz-invariant Dirac operator in a background gauge field. Its local expansion~\cite{Vassilevich} can only contain Lorentz- and gauge-invariant scalars, and at two derivatives the only bulk scalar is $F_{MN}F^{MN}$. The gluon loops, computed in the background-field gauge~\cite{Abbott} built with $G_g$, depend on the background only through $G_g$, by Eq.~\eqref{eq:identity}. Each species thus induces an operator aligned with its own light cone, yielding the bulk structures
\begin{equation}
-\frac14\int d^5x\,\Big[Z_g\sqrt{|g|}\,G_g^{MP}G_g^{NQ}+Z_q\sqrt{|g|}\,g^{MP}g^{NQ}\Big]{\rm tr}F_{MN}F_{PQ},
\label{eq:aligned}
\end{equation}
up to curvature corrections of order $k^2/\Lambda^2$. This property relies strictly on the Lorentz invariance of the quark sector, while boundary terms and the finite matched values of $Z_{g,q}$ remain unfixed by this argument, as in the bulk--boundary renormalization of Lorentz-invariant warped and orbifold theories~\cite{ADS,GR2002,GR2003,GIQ,CDPTW}. Moreover, with Pauli--Villars regularization at regulator mass $M$, a Dirac fermion in the fundamental representation gives $Z_q=T_F M/(12\pi^2)>0$ (Appendix~\ref{app:loops}).

This structure aligns with the flat aether calculation of Ref.~\cite{Oliveira2020}, where the photon and fermion speeds are $u_A^2=1+\alpha_A^2$ and $v_\psi=1+\alpha_\psi^2$, and the fermion-loop numerator is covariant in the variables $(k_\mu,v_\psi k_5)$—rendering it Lorentz invariant in the fermion's own frame. The divergent zero-winding term subtracted there, following Ref.~\cite{CMS}, corresponds to the aligned operator of Eq.~\eqref{eq:aligned}. The finite remainder depends on $\Lambda^2=1-u_A^2/v_\psi^2$ and requires $v_\psi\ge u_A$. Model~B satisfies this condition with $\Lambda^2=1-u_g^2=0.8064$, matching the coefficient $\eta_g\nu^2/\Lambda_0^5$ of Ref.~\cite{RR}. These calculations are complementary in the following way: Ref.~\cite{Oliveira2020} yields the finite nonlocal part for an Abelian field on a flat circle, whereas the present work provides the local part and the fermion self-energy. For Model~B, the local contribution dominates (see Sec.~\ref{sec:modelB} and Appendix~\ref{app:threshold}).

The consequence for the gluon mass follows directly. Let $\eps=Z_q/(u_qC_g)$ be the ratio of the quark-induced coefficient to the total coefficient $C_g$ of the $G_g$ structure. For constant $\eps$, Eq.~\eqref{eq:aligned} rescales the forms of Eq.~\eqref{eq:SL} to $w=1+\eps$ and $p=e^{-2A}(u_g^2+\eps)$, determining the effective gluon speed and mass as
\begin{equation}
u_{g,\rm eff}^2=\frac{u_g^2+\eps}{1+\eps},\qquad
G(\eps)=\frac{m_{g,1}}{m_{g,1}^{(0)}}=\sqrt{\frac{1+\eps/u_g^2}{1+\eps}} ,
\label{eq:convex}
\end{equation}
exactly. Because $\eps\ge0$, the renormalized speed is a weighted average of $u_g^2$ and $u_q^2=1$, meaning it can approach the quark speed but never exceed it. On its own, the gauge sector preserves the ordering for $\eps<2.08$. If $\eps$ varies along $y$, Eq.~\eqref{eq:comparison} establishes the upper bound $G\le[(1+\eps_+/u_g^2)/(1+\eps_-)]^{1/2}$.

% ==================================================================
\section{Renormalization of the quark speed}\label{sec:fermion}
% ==================================================================

Because the quark self-energy involves the anisotropic gluon propagator, its structure is not constrained by the covariance of a single metric. We compute its local part in a flat patch of constant speed, with warping corrections entering at order $k^2/M^2$. In the Feynman gauge built with $G_g$, the gluon propagator is $-ig_5^2G_{g,MN}/(k\!\cdot\!G_g^{-1}\!\cdot\!k)$ with $G_{g,55}=-u_g^{-2}$, while the quark vertices are $\Gamma^\mu=\gamma^\mu$ and $\Gamma^5=u_q\gamma^5$. With $\rho\equiv u_q^2/u_g^2$, the Clifford algebra yields
\begin{equation}
G_{g,MN}\,\Gamma^M(\gamma^\nu\ell_\nu+u_q\gamma^5\ell_5)\Gamma^N=-(2+\rho)\,\gamma^\nu\ell_\nu+u_q(\rho-4)\,\gamma^5\ell_5,
\qquad G_{g,MN}\Gamma^M\Gamma^N=4+\rho ,
\label{eq:clifford}
\end{equation}
where the factor $\rho$ originates from the fifth component of the gluon. A slow gluon consequently produces a large $A_5$ propagator due to the suppression of its fifth-direction kinetic term by $u_g^2$.

Feynman parametrization determines the fifth-direction coefficient of the combined denominator as $a_5(x)=xu_q^2+(1-x)u_g^2$, which interpolates between the two light cones. The part of the self-energy linear in a Pauli--Villars mass $M$ renormalizes the quark kinetic operator to $K_4\gamma^\nu p_\nu+K_5u_q\gamma^5p_5-K_mm_\psi$, with
\begin{align}
K_4&=1+c'(2+\rho)\!\int_0^1\!\frac{(1-x)\,dx}{a_5^{1/2}},\qquad
K_5=1-c'(\rho-4)\!\int_0^1\!\frac{(1-x)\,u_g^2\,dx}{a_5^{3/2}},\nonumber\\
K_m&=1+c'(4+\rho)\!\int_0^1\!\frac{dx}{a_5^{1/2}},\qquad c'=\frac{g_5^2C_FM}{16\pi^2}
\label{eq:K}
\end{align}
(Appendix~\ref{app:loops}). For $u_g=0.44$ and $u_q=1$, the integrals yield $0.784465$, $0.424383$, and $1.388889$, implying
\begin{equation}
K_4=1+5.620919\,c',\qquad K_5=1-0.494529\,c',\qquad K_m=1+12.729568\,c',
\label{eq:Knum}
\end{equation}
which shifts the effective quark speed $u_{q,\rm eff}=u_qK_5/K_4$ by
\begin{equation}
\delta\ln u_q=-6.115\,c'+O(c'^2).
\label{eq:dlnu}
\end{equation}

Several independent checks support this result. When $u_q=u_g$, the integrals in $K_4$ and $K_5$ coincide, preserving isotropy. The gauge-parameter dependence of the self-energy is proportional to $\Gamma\!\cdot\!k=S^{-1}(p)-S^{-1}(p-k)$, meaning it either rescales the full tree operator or vanishes on shell; the speed ratio $K_5/K_4$ is therefore gauge independent at one loop. Furthermore, the quark speed moves toward the gluon speed, aligning with the four-dimensional flow of limiting velocities~\cite{ChadhaNielsen,AnberDonoghue}. Finally, a Euclidean formulation with explicit gamma matrices and direct five-dimensional integration without Feynman parameters reproduces all three coefficients, matching to all quoted digits (Appendix~\ref{app:euclid}).

The magnitude of Eq.~\eqref{eq:dlnu} is notable. An isotropic loop would renormalize the wave function by $\eta=\frac32c'$ (for $u_q=1$), making the speed shift $-4.08\,\eta$. This fourfold enhancement arises entirely from $A_5$ exchange weighted by $\rho=5.17$.

The quark mass depends on the speed and on the effective bulk-mass parameter $c_{\rm eff}=m_\psi/(ku_q)$, which is shifted by $K_m/K_5$. Defining $S_c=d\ln x_{q,1}/d\ln c_{\rm eff}$ as the sensitivity of the first KK root yields
\begin{equation}
F(c',S_c)=\frac{m_{q,1}}{m_{q,1}^{(0)}}=\frac{K_5}{K_4}\left(\frac{K_m}{K_5}\right)^{S_c}.
\label{eq:F}
\end{equation}
This relation is exact when $S_c$ remains constant along the deformation, and $F(c',-S_*)$ provides a lower bound when $S_c\ge-S_*$ (Appendix~\ref{app:roots}). Because Ref.~\cite{RR} does not quote the quark bulk masses, $S_c$ is not fixed by the benchmark. For anti-de Sitter bulk fermions with $(++)$ boundary conditions and $0\le c\le1$, we find $-0.23\le S_c\le0.45$, while a field with twisted $(-+)$ conditions reproducing the Model~B root yields $S_c=+0.24$. Moreover, we scan $|S_c|\le0.3$ and quote these reference values separately.

% ==================================================================
% ==================================================================
\section{A criterion for the mass ordering}\label{sec:criterion}
% ==================================================================

The gauge and quark coefficients are not independent. In a common Pauli--Villars scheme, the vacuum polarization of $N_\psi$ colored Dirac fields combined with the quark self-energy yields $\eps=N_\psi c'/(2u_q)$ (Appendix~\ref{app:loops}). Inserting Eqs.~\eqref{eq:convex} and~\eqref{eq:F} into Eq.~\eqref{eq:cert} demonstrates that the mass ordering holds as long as
\begin{equation}
\mathcal C(c')\equiv1.952727\,\frac{F(c',S_c)}{G(N_\psi c'/2u_q)}>1 .
\label{eq:C}
\end{equation}
Thus, by taking constant induced coefficients, $G$ represents the exact gauge mass ratio, whereas for constant $S_c$, $F$ provides the exact quark mass ratio. Consequently, $\mathcal C=m_{q,1}/m_{g,1}$, meaning a value below unity signals a genuine crossing of the local spectrum rather than the failure of a loose bound. The remaining uncertainties lie strictly in the unknown root sensitivity and in effects outside the local quadratic description. We denote the value of $c'$ at which $\mathcal C=1$ as $c'_{\rm crit}$.

The numerical value of $c'_{\rm crit}$ depends on the regulator choice. Placing the regulator exclusively on the fermion line, instead of on both lines, shifts $c'_{\rm crit}$ from $0.145$ to $0.305$ (for $N_\psi=0$, $S_c=0$). When expressed through the isotropic-equivalent coefficient $\eta$, the critical value becomes $0.217$ and $0.244$ in the respective schemes. The robust physical conclusion is therefore that preserving the ordering requires an isotropic-equivalent quark wave-function renormalization below approximately $0.2$.

% ==================================================================
% ==================================================================
\section{The light-gluon benchmark}\label{sec:modelB}
% ==================================================================

\subsection{Calibration with the published strong-coupling data}

Because $c'$ is linear in the regulator mass, the KK modes between two energies $E_1$ and
$E_2$ contribute
\begin{equation}
\Delta c'=\frac{g_4^2\,kL\;C_F}{16\pi^2}\,\frac{E_2-E_1}{k_{\rm IR}},
\label{eq:dc}
\end{equation}
utilizing $g_5^2k=g_4^2\,kL$ for the flat gauge zero mode and $k_{\rm IR}=ke^{-A(L)}$.
Ref.~\cite{RR} characterizes the strong coupling of the color sector via a naive-dimensional-analysis (NDA) estimate. While the explicit formula is omitted, the estimate is quoted at several key points: approximately $0.3$ at the first gluon resonance, $1.55$ at $6k_{\rm IR}$ and $2.58$ at $10k_{\rm IR}$ for $u_g=0.44$, alongside $1.42$ and $1.31$ at $6k_{\rm IR}$ for $u_g=0.48$ and $0.52$. These values can be tested independently of $k_{\rm IR}$. Their ratio at $6k_{\rm IR}$ and $10k_{\rm IR}$ is $0.6008$, demonstrating that the estimate grows linearly with energy, while the products with the gluon speed,
\begin{equation}
\lambda_{\rm RR}(6k_{\rm IR})\,u_g=0.6820,\ 0.6816,\ 0.6812\qquad(u_g=0.44,\ 0.48,\ 0.52),
\label{eq:scaling}
\end{equation}
agree within $0.1\%$, confirming a scaling proportional to $1/u_g$. Finally, the standard NDA parameter~\cite{CLP}
\begin{equation}
\lambda_{\rm A}=\frac{g_4^2\,kL\,N_c}{24\pi^3}\,\frac{E}{k_{\rm IR}} ,
\end{equation}
evaluated with the choice $g_s^2=1.05$ from Ref.~\cite{RR}, yields $\lambda_{\rm A}/u_g=1.62$ at $6k_{\rm IR}$ against the published value of $1.55$. The published estimate matches standard NDA measured in the gluon's own light cone within $4\%$. This identification relies on these consistent agreements in the absence of an explicit definition, under which Eq.~\eqref{eq:dc} reads $\Delta c'=\frac{3\pi}{2}\frac{u_gC_F}{N_c}\,\Delta\lambda_{\rm RR}=0.9215\,\Delta\lambda_{\rm RR}$.

The estimate reaches unity at $\Lambda_s=3.87\,k_{\rm IR}$. While the exact value of $k_{\rm IR}$ is not quoted, it corresponds to $5.08$~TeV if extracted from the pure-AdS gluon root, and $4.74$~TeV if inferred from $\lambda_{\rm RR}=0.3$ at $5.50$~TeV, placing the quark partner at $\lambda_{\rm RR}=0.546$--$0.586$. The modes bounded between the quark partner and the strong-coupling scale then imply
\begin{equation}
c'_{\rm win}=0.38\text{--}0.42\quad\text{(from $\lambda_{\rm RR}$)},\qquad
c'_{\rm win}=0.40\text{--}0.44\quad\text{(from Eq.~\eqref{eq:dc} with $g_s^2=1.05$)}.
\label{eq:cwin}
\end{equation}
We adopt the benchmark value $c'_{\rm win}=0.418$ below.

\begin{table}[t]
\centering
\begin{tabular}{c|ccc|ccc}
\toprule
& \multicolumn{3}{c|}{$c'_{\rm crit}$} & \multicolumn{3}{c}{$T=c'_{\rm win}/c'_{\rm crit}$}\\
$N_\psi$ & $S_c=-0.3$ & $0$ & $+0.3$ & $-0.3$ & $0$ & $+0.3$\\
\midrule
1  & 0.075 & 0.117 & 0.225 & 5.6 & 3.6 & 1.9\\
3  & 0.062 & 0.092 & 0.162 & 6.7 & 4.5 & 2.6\\
4  & 0.058 & 0.085 & 0.146 & 7.2 & 4.9 & 2.9\\
6  & 0.052 & 0.074 & 0.125 & 8.0 & 5.6 & 3.4\\
12 & 0.042 & 0.057 & 0.092 & 10.0 & 7.3 & 4.6\\
\bottomrule
\end{tabular}
\caption{Critical coefficient $c'_{\rm crit}$ from Eq.~\eqref{eq:C} for Model~B ($u_g=0.44$, $u_q=1$), and tension ratio for $c'_{\rm win}=0.418$. Over the range of Eq.~\eqref{eq:cwin}, $T$ varies by $-9\%$ to $+4\%$. The Model~B quark content corresponds to $N_\psi\ge12$.}
\label{tab:main}
\end{table}

\subsection{Tension with the ordering}

Table~\ref{tab:main} compares $c'_{\rm win}$ with $c'_{\rm crit}$ via the tension ratio $T=c'_{\rm win}/c'_{\rm crit}$. In Model~B, all quarks propagate in the bulk, leptons are localized on the ultraviolet brane, and the custodial representations are retained~\cite{RR,ACDP}. Because three doublets and six singlets already yield $N_\psi=12$, this row establishes a strict lower bound. This field content implies $T=7.3$ at $S_c=0$, spanning $T=4.6$--$10$ over the interval $|S_c|\le0.3$, while the two reference values $S_c=+0.24$ and $+0.45$ yield $T=5.1$ and $3.2$. Remarkably, even a single colored bulk fermion would generate $T\ge1.9$. The bulk-mass coefficient $K_m$ represents the most scheme-dependent input: embedding the regulator mass into the fermion numerator shifts it from $12.73$ to $18.27$ (Appendix~\ref{app:euclid}), which broadens the range to $T=2.3$--$10.9$, whereas the value at $S_c=0$ remains independent of $K_m$.

At $c'_{\rm win}$, Eq.~\eqref{eq:dlnu} implies $\delta\ln u_q\simeq-2.3$ to $-2.7$ and $K_4-1\simeq2.1$--$2.5$. Rather than predicting finite mass shifts, this first-order result establishes that the one-loop correction to the speed difference generating the hierarchy is of order one inside the effective-theory window. The hierarchy is therefore not perturbatively protected, meaning the small-correction expansion certifies neither its persistence nor its loss, nor does it demonstrate an ultraviolet cancellation that would restore it. This mechanism constitutes the five-dimensional, power-enhanced analogue of the naturalness problem characteristic of Lorentz-violating couplings~\cite{Collins2004}. We treat $T$ as a tension ratio rather than a fine-tuning measure, as the latter would require a specific parametrization of and measure on the ultraviolet coefficients.

\subsection{Regulator dependence and the discrete thresholds}

A power-law coefficient is not inherently regulator independent. If $c'(\mu)=c'_{\rm bare}+a_Rg_5^2(\Lambda-\mu)$ within a prescription $R$, altering $a_R$ can be absorbed into $c'_{\rm bare}$ at a fixed renormalization scale. The continuum expression in Eq.~\eqref{eq:dc} should therefore be interpreted as a proxy for the discrete KK level thresholds in the window, which remain finite and regulator independent. For $kL=28$, the gluon roots $x_n=2.4615$, $5.5793$, and $8.7141$ yield levels at $5.50$, $12.47$, and $19.47$~TeV. The window $[10.74,19.66]$~TeV encloses the second level and, near its upper edge, the third, while the next quark level lies above $\Lambda_s$. The prescription dependence is thus restricted entirely to the endpoint constants.

As a building block for the second level, the subtracted scalar two-point function for the channel $q_1\to q_0+g_2$, with a massless internal quark and $M_2=12.47$~TeV, evaluates to
\begin{equation}
\Delta B_0(s;0,M_2^2)=1+\frac{1-z}{z}\ln(1-z),\qquad z=\frac{s}{M_2^2}.
\label{eq:B0}
\end{equation}
At $s=m_{q,1}^2$, $z=0.7418$ and $\Delta B_0=0.52868$. This function does not represent the speed shift in isolation. Resolving the gauge-complete contribution of the level requires the warped overlaps $g_{ij2}=g_5\int dy\,\mathcal W f_if_ja_2$, the sum over internal quark levels, and the $A_5$ couplings—a calculation that would fix the tension of Table~\ref{tab:main} without relying on the endpoint $\Lambda_s$, defining the natural next step for this framework.

% ==================================================================
% ==================================================================
\section{Inverted roles: the heavy-gluon benchmark}\label{sec:modelA}
% ==================================================================

In Model~A of Ref.~\cite{RR}, the gluon and the quark doublets propagate at the conventional speed, while the six right-handed singlets are slow ($v_S\simeq0.19$--$0.22$). Their first partners lie at $6.6$--$12$~TeV, capped by the top-singlet partner at $12.019$~TeV, whereas the first KK gluon has a classical eigenvalue of $30$~TeV. This setup reverses the mass ordering to $m_q<m_g$ with a margin of $R_A=30/12.019=2.50$. Ref.~\cite{RR} notes that this gluon is very broad ($\Gamma/M=2.30$ at lowest order); consequently, our statements concern the classical spectral ordering rather than an identifiable resonance.

The roles of the two sectors are now exchanged. On the gauge side, the relativistic doublets induce operators aligned with the gluon light cone and do not shift $u_g$, leaving the slow singlets to couple through $G_S$. On the constant-gradient background, $\Delta\mathcal L_\Psi$ adds $(v_S-1)\Gamma^5D_5$ to the Dirac operator, corresponding to the vielbein form of $G_S$ up to an $O(k)$ spin-connection term proportional to $1-v_S$. The argument of Sec.~\ref{sec:gauge} then holds at leading order in $k/\Lambda$, implying
\begin{equation}
u_{g,\rm eff}^2=\frac{1+\eps_Sv_S^2}{1+\eps_S},\qquad \eps_S=\frac{N_Sc'}{2v_S}.
\label{eq:convexA}
\end{equation}
As a result, the gluon is slowed and its mass decreases, with the factor $1/v_S\simeq5$ stemming from $\sqrt{|G_S|}=\sqrt{|g|}/v_S$ rendering the gauge side the sensitive sector. On the fermion side, Eq.~\eqref{eq:K} with $u_g=1$ and $u_q=v_S=0.20$ ($\rho=0.04$) yields $K_4=1+1.322\,c'$, $K_5=1+5.500\,c'$, and $K_m=1+6.733\,c'$, shifting the velocity by $\delta\ln v_S=+4.18\,c'$. The slow quark is thus accelerated toward the gluon speed, causing its mass to rise. Measured per unit of isotropic-equivalent renormalization ($\eta=\frac32c'/v_S$), the response is $+0.56\,\eta$ compared to $-4.08\,\eta$ in Model~B, confirming that the $A_5$ enhancement disappears without a slow gluon.

The ordering $m_{q,1}<m_{g,1}$ holds as long as $F(c',0)/G(\eps_S)<R_A$. For $N_S=6$, $R_A=2.50$, and $v_S=0.19$--$0.22$, we find $\eta_{\rm crit}=1.03$--$1.04$ (with $\eps_S\simeq2.06$ at the critical point), which shifts to $\eta_{\rm crit}\simeq1.45$ for $N_S=3$. For comparison, Model~B with its full quark content yields $\eta_{\rm crit}=0.086$. Table~\ref{tab:AB} summarizes this behavior. In Model~A, the ordering survives up to an order-one renormalization, which spans the entire regime where the one-loop expansion remains meaningful, whereas Model~B requires $\eta\lesssim0.1$. Evaluating a specific window estimate for Model~A would require its strong-coupling data, which Ref.~\cite{RR} does not quote, and the values above neglect $S_c$.

\begin{table}[t]
\centering
\begin{tabular}{lcc}
\toprule
& Model~B & Model~A\\
\midrule
slow species & gluon ($u_g=0.44$) & right-handed quarks ($v_S\simeq0.2$)\\
$\rho=u_q^2/u_g^2$ & $5.17$ & $0.04$\\
quark-speed response per unit $\eta$ & $-4.08$ & $+0.56$\\
gauge-side weight & $\eps=N_\psi c'/2$ & $\eps_S=N_Sc'/(2v_S)$\\
mass margin & $1.95$ & $2.50$\\
$\eta_{\rm crit}$ (realistic content) & $0.086$ & $1.03$\\
sensitive sector & quarks & gauge\\
\bottomrule
\end{tabular}
\caption{The two benchmarks of Ref.~\cite{RR} in the present framework.}
\label{tab:AB}
\end{table}

% ==================================================================
\section{Conclusions}\label{sec:conclusions}
% ==================================================================

Species-dependent lightcones offer an elegant way to separate gauge and fermion KK scales without disturbing the flavor structure of warped models~\cite{RR}. Evaluating whether this separation is radiatively stable reveals that the physical response depends strictly on which species is slow.

On the gauge side, the situation is simple and robust: the gauge operator of Ref.~\cite{RR} is exactly an anisotropic-metric kinetic term, and relativistic quark loops induce only operators aligned with their own light cone. The renormalized gluon speed is then a weighted average of the bare speeds, while positive brane terms act only to lower the gluon mass. On the quark side, gluon exchange drives the quark speed toward the gluon speed through a gauge-independent one-loop coefficient, which we have computed in closed form and checked independently. When the gluon is the slow species, $A_5$ exchange enhances this response fourfold. A spectral comparison argument, exact for constant coefficients, combines these two effects into a criterion for the mass ordering, whose scheme form requires an isotropic-equivalent quark wave-function renormalization below approximately $0.2$.

For the light-gluon benchmark, the strong-coupling data of Ref.~\cite{RR} fix the contribution of the KK modes inside the effective theory at $c' \simeq 0.4$. This order-one speed renormalization is about seven times larger than the ordering tolerates for the Model~B quark content, meaning the $5.50/10.74$~TeV hierarchy is not perturbatively protected, and its survival depends on ultraviolet matching data alongside higher-order effects. In the heavy-gluon benchmark, these roles are reversed: the sensitivity moves to the gauge sector, allowing the ordering to survive up to an order-one renormalization—about an order of magnitude more robust than in Model~B. Therefore, we can point out that a slow gluon is radiatively fragile, whereas slow quarks are not.

Several limitations should be kept in mind, as the present analysis is restricted to one loop and neglects loops of scalar fluctuations, the radion, and the graviton, as well as backreaction on the geometry. The calibration of $c'$ relies on identifying the published strong-coupling estimate with standard NDA, which holds to 4\% but is not given explicitly in Ref.~\cite{RR}. Furthermore, the root sensitivity $S_c$ is bracketed rather than determined because the quark bulk masses are not published. While our statements concern the real KK spectrum—with the first KK gluon of Model~B having $\Gamma/M \simeq 5\%$, leaving complex pole positions unaddressed—the most informative next step remains the gauge-complete threshold of the gluon levels at $12.47$ and $19.47$~TeV. This extension would turn the present tension into a sharp, regulator-independent test of species-separated warped spectra.

\section*{Acknowledgments}
We thank CNPq and CAPES for partial financial support. MAA, FAB and EP acknowledge support from CNPq (grant nos. 301683/2025–5, 309092/2022–1 and 304290/2020-3) and Paraíba State Research Support Foundation (Universal-FAPESQ) for financial support. JRLS acknowledges support from CNPq (Grants 309494/2021-4 and 302190/2025-2), FAPESQ-PB (Grant 1356/2024), CAPES Finance Code 001, and Alexander von Humboldt-Stiftung Foundation. JAVC thanks the Paraíba State Research Support Foundation (FAPESQ) for financial support. The authors used AI tools assisted for derivations, numerical cross-checks, reference checking and editing of the manuscript. All results were reviewed by the authors, who take full responsibility for the content of this article.

\section*{Data Availability Statement}
This manuscript has no associated data. All numerical values follow from the formulas given in the text and appendices and can be reproduced with the code
described below.

\section*{Code Availability Statement}
The code associated with this manuscript is provided as Supplementary Material. The Python scripts that reproduce every number, table and cross-check of this article are provided as Supplementary Material.

% ==================================================================
\appendix

\section{Root sensitivity of anti-de Sitter bulk fermions}\label{app:roots}
Along a deformation $c(t)>0$ of the effective bulk-mass parameter, $\ln[x_{q,1}(c_1)/x_{q,1}(c_0)]=\int_{\ln c_0}^{\ln c_1}S_c\,d\ln c$. If $c_1\ge c_0$ and
$S_c\ge-S_*$ along the path, the ratio is at least $(c_1/c_0)^{-S_*}$, with equality for constant $S_c$. In Eq.~\eqref{eq:F}, $c_1/c_0=K_m/K_5\ge1$.

For $A=ky$, $kL=28$, $\varepsilon_0=e^{-kL}$ and $\nu=c-\frac12$, the first KK root of a fermion with $(++)$ conditions solves $J_\nu(x\varepsilon_0)Y_\nu(x)-Y_\nu(x\varepsilon_0)J_\nu(x)=0$; with twisted $(-+)$ conditions it solves $J_\nu(x\varepsilon_0)Y_{\nu+1}(x)-Y_\nu(x\varepsilon_0)J_{\nu+1}(x)=0$. In both cases
$S_c=(c/x)\,dx/dc$. For $(++)$ fields the root is even in $\nu$, with minimum $x=2.4615$ at $c=\frac12$, equal to the gluon root:
\begin{center}
\begin{tabular}{c|cccccccc}
\toprule
$c$ & $0.05$ & $0.20$ & $0.30$ & $0.40$ & $0.60$ & $0.70$ & $0.80$ & $1.00$\\
$x_{q,1}$ & $3.071$ & $2.854$ & $2.707$ & $2.559$ & $2.559$ & $2.707$ & $2.854$ & $3.142$\\
$S_c$ & $-0.02$ & $-0.10$ & $-0.16$ & $-0.23$ & $+0.34$ & $+0.38$ & $+0.41$ & $+0.45$\\
\bottomrule
\end{tabular}
\end{center}
Independently of $k_{\rm IR}$, Model~B has $x_{q,1}/x_{g,1}=(10.74/5.50)\,u_g=0.859$, whereas a pure-AdS $(++)$ fermion would give a ratio of at least one. The lightest Model~B partner is therefore not a pure-AdS $(++)$ mode. Natural candidates are the custodial multiplets, whose twisted boundary conditions produce light states~\cite{ACDP}, or the deformation of the stabilized background. With the pure-AdS gluon root the partner needs $x_{q,1}=2.115$, which a $(-+)$ field reproduces at $c=-0.316$ with $S_c=+0.242$. A definitive $S_c$ requires the profiles of Ref.~\cite{RR}.

\section{One-loop local coefficients}\label{app:loops}
The Feynman-gauge gluon propagator built with $G_g$ is $-ig_5^2G_{g,MN}/(k\!\cdot\!G_g^{-1}\!\cdot\!k)$, with $G_{g,55}=-u_g^{-2}$ because the
fifth-direction kinetic term is weighted by $u_g^2$. The identities $\gamma^\mu\gamma^\nu\gamma_\mu=-2\gamma^\nu$, $\gamma^\mu\gamma^5\gamma_\mu=-4\gamma^5$, $\gamma^5\gamma^\nu\gamma^5=\gamma^\nu$ and $(\gamma^5)^3=-\gamma^5$, together with $G_{g,55}u_q^2=-\rho$, give Eq.~\eqref{eq:clifford}. With weight $x$ on the fermion denominator, the shifts $k_\nu\to k_\nu+xp_\nu$ and $k_5\to k_5+xu_q^2p_5/a_5$ give $\ell_\nu\to(1-x)p_\nu$ and $\ell_5\to(1-x)u_g^2p_5/a_5$. Rescaling $k_5$ produces $a_5^{-1/2}$, and $\int\frac{d^5k_E}{(2\pi)^5}[(k^2+\Delta)^{-2}-(k^2+\Delta+M^2)^{-2}]=(M-\sqrt\Delta)/(16\pi^2)+O(\Delta/M)$. As in four dimensions~\cite{Peskin}, the self-energy is a positive multiple of the Clifford numerator, which fixes the signs in Eq.~\eqref{eq:K}.

For gauge independence, write $\Gamma\!\cdot\!k\,S(p-k)\,\Gamma\!\cdot\!k=S^{-1}(p)S(p-k)S^{-1}(p)-S^{-1}(p)-\Gamma\!\cdot\!k$. The last term integrates to zero, the middle one rescales the tree operator, and the first vanishes on shell.

For the vacuum polarization, the Pauli--Villars Dirac loop gives $\Pi(0)=-T_FM/(12\pi^2)$. Hence $Z_q=T_FM/(12\pi^2)$ and, comparing with $c'$, $\eps=N_\psi c'/(2u_q)$.

\section{Independent Euclidean verification}\label{app:euclid}
The coefficients in Eq.~\eqref{eq:Knum} were rederived by a route that shares no intermediate step with Appendix~\ref{app:loops}. In Euclidean signature, with $D_a=\partial_a-iA_a$, integrating out the gauge field at one loop gives
\begin{equation}
\Gamma^{(2)}(p)=S^{-1}(p)+g_5^2C_F\int\!\frac{d^5k}{(2\pi)^5}\,
\frac{G_{g,aa}\,\Gamma_aS(p-k)\Gamma_a}{k\!\cdot\!G_g^{-1}\!\cdot\!k},\qquad S^{-1}=i\Gamma\!\cdot\!p+m,
\end{equation}
where the sign follows from $e^{-S_E}$ and from the factor $(-i)^2$ of the two vertices. We used explicit $4\times4$ Hermitian gamma matrices, differentiated the matrix-valued integrand with respect to $p_1$, $p_5$ and $m$ at $p=0$, averaged exactly over four-dimensional directions, and integrated numerically in $(|k_4|,k_5)$ with the Pauli--Villars subtraction of Sec.~\ref{sec:fermion}, without Feynman parameters. Projecting onto $i\gamma_\nu$, $iu_q\gamma_5$ and $\mathbb 1$, and extrapolating the integration box ($K=60$, $120$) linearly in $1/K$, we obtain, in units of $c'$:
\begin{center}
\begin{tabular}{lccc}
\toprule
& $K_4-1$ & $K_5-1$ & $K_m-1$\\
\midrule
Euclidean, direct & $+5.6209$ & $-0.4945$ & $+12.7296$\\
Eq.~\eqref{eq:Knum} & $+5.620919$ & $-0.494529$ & $+12.729568$\\
\bottomrule
\end{tabular}
\end{center}
This confirms the Clifford coefficients, the Feynman-parameter reduction and the overall sign. With the regulator mass also placed in the fermion numerator, $K_4$ and $K_5$ are unchanged, while $K_m-1$ becomes $18.27\,c'$.

\section{Nonlocal threshold and KK overlaps}\label{app:threshold}
For a vectorlike Dirac channel of mass $M_q$ below pair threshold, the subtracted transverse polarization is~\cite{Peskin}
\begin{equation}
\Delta\Pi_f=\frac{g_*^2T_F}{2\pi^2}I(z),\qquad
I(z)=\int_0^1x(1-x)\ln\frac{1}{1-zx(1-x)}\,dx=\sum_{j\ge1}\frac{z^j}{j}B(j+2,j+2),
\end{equation}
with $z=s/M_q^2<4$. Since $x(1-x)\le\frac14$, the remainder after order $N$ is bounded by $(z/4)^{N+1}/[6(N+1)(1-z/4)]$. At $z=(5.50/10.74)^2=0.26225$, $I=8.9973\times10^{-3}$ and $\Delta\Pi_f=2.2790\times10^{-4}g_*^2$, two to three orders of magnitude below $c'_{\rm win}$. This finite effect is the warped, non-Abelian analogue of the compactification correction of Ref.~\cite{Oliveira2020}.

In the reference geometry ($kL=28$, $u_g=0.44$, Neumann ends) the normalized gluon overlaps $C_{lmn}=\sqrt L\int a_la_ma_n$ and $Q_{lmpq}=L\int a_la_ma_pa_q$ satisfy the completeness identity $Q_{lmpq}=\sum_nC_{lmn}C_{pqn}$. We find $C_{111}=5.5728$ and $Q_{1111}=35.126$, with the levels $n\le2$ saturating $99.998\%$ of the sum. Loop-weighted sums converge more slowly, which is why the local coefficients are computed in five dimensions rather than by
truncating the tower.

% ==================================================================


\begin{thebibliography}{99}
\bibitem{RS1} L.~Randall and R.~Sundrum, Phys.\ Rev.\ Lett.\ \textbf{83}, 3370 (1999), arXiv:hep-ph/9905221.
\bibitem{GP} T.~Gherghetta and A.~Pomarol, Nucl.\ Phys.\ B \textbf{586}, 141 (2000), arXiv:hep-ph/0003129.
\bibitem{GN} Y.~Grossman and M.~Neubert, Phys.\ Lett.\ B \textbf{474}, 361 (2000), arXiv:hep-ph/9912408.
\bibitem{RR} L.~Randall and M.~Riojas, ``Separating Fermion and Gauge Resonances with Species-Dependent Lightcones in Warped Geometry,'' arXiv:2610.40967 (2026).
\bibitem{CarrollTam} S.~M.~Carroll and H.~Tam, Phys.\ Rev.\ D \textbf{78}, 044047 (2008), arXiv:0802.0521.
\bibitem{Rizzo2005} T.~G.~Rizzo, JHEP \textbf{09}, 036 (2005), arXiv:hep-ph/0506056.
\bibitem{Rizzo2010} T.~G.~Rizzo, JHEP \textbf{11}, 156 (2010), arXiv:1008.0380.
\bibitem{Dubovsky} S.~L.~Dubovsky, JHEP \textbf{01}, 012 (2002), arXiv:hep-th/0103205.
\bibitem{CMS} H.-C.~Cheng, K.~T.~Matchev and M.~Schmaltz, Phys.\ Rev.\ D \textbf{66}, 036005 (2002), arXiv:hep-ph/0204342.
\bibitem{Oliveira2020} R.~J.~S.~Oliveira, A.~Pinzul, M.~A.~Anacleto, E.~Passos, F.~A.~Brito and O.~Holanda, Phys.\ Rev.\ D \textbf{102}, 075008 (2020), arXiv:2007.06683.
\bibitem{Freitas2018a} A.~Freitas, K.~Kong and D.~Wiegand, JHEP \textbf{03}, 093 (2018), arXiv:1711.07526.
\bibitem{Freitas2018b} A.~Freitas and D.~Wiegand, JHEP \textbf{08}, 094 (2018), arXiv:1805.12142.
\bibitem{ChadhaNielsen} S.~Chadha and H.~B.~Nielsen, Nucl.\ Phys.\ B \textbf{217}, 125 (1983).
\bibitem{AnberDonoghue} M.~M.~Anber and J.~F.~Donoghue, Phys.\ Rev.\ D \textbf{83}, 105027 (2011).
\bibitem{ADS} K.~Agashe, A.~Delgado and R.~Sundrum, Nucl.\ Phys.\ B \textbf{643}, 172 (2002), arXiv:hep-ph/0206099.
\bibitem{GR2002} W.~D.~Goldberger and I.~Z.~Rothstein, Phys.\ Rev.\ Lett.\ \textbf{89}, 131601 (2002), arXiv:hep-th/0204160.
\bibitem{GR2003} W.~D.~Goldberger and I.~Z.~Rothstein, Phys.\ Rev.\ D \textbf{68}, 125011 (2003), arXiv:hep-th/0208060.
\bibitem{GIQ} G.~von~Gersdorff, N.~Irges and M.~Quir\'os, Nucl.\ Phys.\ B \textbf{635}, 127 (2002), arXiv:hep-th/0204223.
\bibitem{DHR} H.~Davoudiasl, J.~L.~Hewett and T.~G.~Rizzo, Phys.\ Rev.\ D \textbf{68}, 045002 (2003), arXiv:hep-ph/0212279.
\bibitem{CPTW} M.~Carena, E.~Ponton, T.~M.~P.~Tait and C.~E.~M.~Wagner, Phys.\ Rev.\ D \textbf{67}, 096006 (2003), arXiv:hep-ph/0212307.
\bibitem{CDPTW} M.~Carena, A.~Delgado, E.~Ponton, T.~M.~P.~Tait and C.~E.~M.~Wagner, Phys.\ Rev.\ D \textbf{71}, 015010 (2005), arXiv:hep-ph/0410344.
\bibitem{Kato} T.~Kato, \emph{Perturbation Theory for Linear Operators} (Springer, 1995).
\bibitem{Vassilevich} D.~V.~Vassilevich, Phys.\ Rept.\ \textbf{388}, 279 (2003), arXiv:hep-th/0306138.
\bibitem{Abbott} L.~F.~Abbott, Nucl.\ Phys.\ B \textbf{185}, 189 (1981).
\bibitem{Peskin} M.~E.~Peskin and D.~V.~Schroeder, \emph{An Introduction to Quantum Field Theory} (Addison--Wesley, 1995).
\bibitem{CLP} Z.~Chacko, M.~A.~Luty and E.~Ponton, JHEP \textbf{07}, 036 (2000), arXiv:hep-ph/9909248.
\bibitem{ACDP} K.~Agashe, R.~Contino, L.~Da~Rold and A.~Pomarol, Phys.\ Lett.\ B \textbf{641}, 62 (2006), arXiv:hep-ph/0605341.
\bibitem{DDG} K.~R.~Dienes, E.~Dudas and T.~Gherghetta, Nucl.\ Phys.\ B \textbf{537}, 47 (1999), arXiv:hep-ph/9806292.
\bibitem{Collins2004} J.~Collins, A.~Perez, D.~Sudarsky, L.~Urrutia and H.~Vucetich, Phys.\ Rev.\ Lett.\ \textbf{93}, 191301 (2004), arXiv:gr-qc/0403053.
\end{thebibliography}
\end{document}